\documentclass[twocolumn,tighten]{aastex61}
\usepackage{amssymb, amsmath,framed}

\usepackage{bm}
\expandafter\ifx\csname package@font\endcsname\relax\else
 \expandafter\expandafter
 \expandafter\usepackage
 \expandafter\expandafter
 \expandafter{\csname package@font\endcsname}
\fi
\def\bq{\begin{equation}}
\def\eq{\end{equation}}
\def\bqy{\begin{eqnarray}}
\def\eqy{\end{eqnarray}}

\begin{document}
\title{Risks for life on habitable planets from superflares of their host stars}

\correspondingauthor{Manasvi Lingam}
\email{manasvi.lingam@cfa.harvard.edu}

\author{Manasvi Lingam}
\affiliation{Harvard-Smithsonian Center for Astrophysics, 60 Garden St, Cambridge, MA 02138, USA}
\affiliation{John A. Paulson School of Engineering and Applied Sciences, Harvard University, 29 Oxford St, Cambridge, MA 02138, USA}
\author{Abraham Loeb}
\affiliation{Harvard-Smithsonian Center for Astrophysics, 60 Garden St, Cambridge, MA 02138, USA}

\begin{abstract}
We explore some of the ramifications arising from superflares on the evolutionary history of Earth, other planets in the Solar system, and exoplanets. We propose that the most powerful superflares can serve as plausible drivers of extinction events, and that their periodicity could correspond to certain patterns in the terrestrial fossil diversity record. On the other hand, weaker superflares may play a positive role in enabling the origin of life through the formation of key organic compounds. Superflares could also prove to be quite detrimental to the evolution of complex life on present-day Mars and exoplanets in the habitable zone of M- and K-dwarfs. We conclude that the risk posed by superflares has not been sufficiently appreciated, and that humanity might potentially witness a superflare event in the next $\sim 10^3$ years leading to devastating economic and technological losses. In light of the many uncertainties and assumptions associated with our analysis, we recommend that these results should be viewed with due caution.
\end{abstract}

\section{Introduction}
Flares are eruptions of high-energy radiation from stars, and phenomena associated with these events have been recorded, and studied, throughout human history \citep{Eat80,Vaq07}. One of the powerful solar flares on record, the Carrington event, dates back to more than $150$ years ago \citep{Carr59}. Solar (and stellar) flares have been extensively studied in recent times for a multitude of reasons. There has been a great deal of interest in understanding the physical mechanisms responsible for their origin, usually through magnetic reconnection \citep{Pri14,Jan17} resulting in the rapid release of magnetic energy \citep{ShiMa11,JAD15,CLHB,CLHB17}. In addition, flares have been exhaustively studied in the context of space weather predictions \citep{Schw06,Barn11}, as they can indirectly cause damage to satellites and astronauts in orbit. As stellar flares are typically associated with the emission of ultraviolet (UV) radiation and high-energy protons, several studies have been undertaken to gauge the robustness of life on Earth as well as other exoplanets to these events \citep{Rind02,Buc06,MT11,Dart11,AM14}.

A common theme in most of these papers is that the flares studied were not particularly extreme, as most of them were characterized by energies $\lesssim 10^{32}$ erg. However, the launch of the \emph{Kepler} mission to detect exoplanets greatly altered, and advanced, our understanding of the statistics of flares \citep{Wal11,Mae12,Shi13,Cand14}. The analysis of the \emph{Kepler} data revealed that highly energetic flares, dubbed superflares, occurr on M-, K- and G-type stars with a fairly high frequency. In turn, this discovery reignited interest in the possibility that superflares could occur on the Sun over the span of a few thousands of years \citep{KS13}. In parallel, based on evidence from radionuclides in tree rings \citep{MNMN}, it was  suggested that the deduced spike in cosmic rays could potentially be explained by a solar superflare that erupted in AD 775 \citep{MT12,Uso13}.

In light of the mounting evidence concerning the importance of superflares, we carry out below an analysis of their implications for life on Earth and exoplanets. The outline of the paper is as follows. In Sec. \ref{SecSync}, we present connections between extinction events in the fossil record and the frequency of superflares and highlight some of the assumptions inherent in our analysis. We follow this by analyzing the effects of large superflares on Earth in Sec. \ref{SecEff}, and conclude that they can trigger mass extinctions. In Sec. \ref{SecImp}, we explore the negative and positive consequences of superflares for life on Mars, Venus and exoplanets orbiting low-mass stars. We also delineate the economic risks posed by superflares to human civilization. Finally, we summarize the salient results of the paper in Sec. \ref{SecConc}.

\section{Connections between the timing of superflares and species extinctions} \label{SecSync}
In this Section, we shall explore the timescales associated with large superflares on the Sun and outline possible connections with the fossil diversity record. We also delineate the caveats and assumptions in our model.

\subsection{Timescales for superflares and mass extinctions}\label{SSecT}
Although large-scale extinction events have occurred multiple times on Earth, their exact number remains uncertain. One of the important hypotheses put forward concerns the existence of putative periodic patterns in the fossil extinction record \citep{Raup86,HW97,Cour99}. The estimates for the periodic timescale have typically ranged from $26$ Myr \citep{RS84,RS86} to $62$ Myr \citep{RM05}. The evidence in favor of and against this periodicity has been explored extensively over the past three decades \citep{PS87,Ben95,Bam06,Al08,MB14}. To explain these extinction events, a wide range of astrophysical phenomena such as gamma ray bursts (GRBs), supernovae, the presence of a distant solar companion, and comet impacts have been invoked; the reader may consult \citet{BJ09} for further details.

If we choose a periodic timescale of $\tau = 26$ Myr and hypothesize that the extinctions are caused by an astrophysical phenomenon, the latter must repeat after this interval of time. We shall posit that solar superflares serve as a driver of these extinction events, and thereby examine whether they constitute a viable mechanism. 

We begin by noting that superflares are extremely rare events, and solar observations have not been undertaken for sufficiently long periods to directly document their existence \citep{Uso17}. Fortunately, the observations of $\sim 10^5$ solar-type stars by the \emph{Kepler} mission have yielded a wealth of data \citep{Mae12}. For slowly rotating, G-type stars like the Sun, the following relation was empirically determined:
\begin{equation} \label{Freq}
    \frac{d N}{d E} \propto E^{-\alpha} \quad \alpha \gtrsim 2,
\end{equation}
where $N$ was the occurrence rate of superflares as a function of the energy $E$ \citep{Mae12}. A detailed analysis of the \emph{Kepler} data led to the conclusion that flares of energy $\sim 10^{34}$ erg would occur every $\sim 2000$ yrs \citep{Shi13}. In this context, we observe that a superflare on the Sun with energy $\sim 10^{34}$ erg that occurred in AD $775$ ($\sim 1250$ years ago) has been posited \citep{MT12,Uso13,Mek15}, although the proposed evidence and reasoning are open to other interpretations \citep{MNMN,CTD14,NN15}. Another energetic event dating from AD 993 has been associated with a superflare \citep{MMN13,Mek15}; however, the corresponding astronomical evidence has been critiqued \citep{Step15}. As the Sun has lower activity levels with respect to most solar-type stars, it has been suggested that the frequency of superflares on the Sun with $\sim 10^{34}$ erg could be lower by an order of magnitude \citep{KK16}. 

We are now in a position to answer the question: what is the energy $E$ of a superflare that occurs with a frequency of $\sim 20$ Myr? Using the above information in conjunction with $\alpha = 2.3$ \citep{Mae12}, we are led to conclude that $E \sim 10^{37}$ erg. This raises the immediate question as to whether flares of this magnitude are achievable on solar-type stars, since the \emph{Kepler} sample only yielded values $\lesssim 2 \times 10^{36}$ erg \citep{Shi13,Mae15}; on the other hand, it must be recognized that the difference in the two maximal values is less than one order of magnitude (a factor of $5$). In order to answer this question, we shall rely upon a combination of empirical and theoretical considerations.

From the observational standpoint, we note that flares with energies $\sim 10^{38}$ erg have been documented in G-type stars \citep{SKD00}. The result is pertinent since these stars are: (i) not rapid rotators, (ii) typically single, and (iii) not very young. Thus, in many respects these stars are similar to the Sun, thereby suggesting that equally large flares may, in principle, also be manifested in the latter. We also wish to point out that superflares with energies $\sim 10^{37}$ erg have been documented for some G-type stars studied by the \emph{Kepler} mission \citep{Wal11,Bas11}. Based on the empirical evidence, it was suggested in Sec. V of \citet{Sch12} that the theoretical upper bound for superflares on ``Sun-like'' stars on the main sequence would be $\sim 10^{37}$ erg. However, we caution that some of these stars possess higher ambient surface magnetic fields than the Sun; consequently, superflares on the Sun might have a very low probability of occurrence as discussed further in Sec. \ref{SSecCav}. 

\citet{KS13} argued that superflares on the Sun can arise provided that sufficiently large sunspots, approximately $30\%$ of the surface area, are formed. Furthermore, some empirical evidence from other Sun-like stars also indicates that superflares could occur on the Sun \citep{Nog14}. If we consider active low-mass stars, a spot-coverage fraction of $\sim 0.4$ appears to be fairly common \citep{JJ13}. The energy $E$ of the flare can be expressed as
\begin{equation} \label{FlareEn}
    E = \epsilon E_\mathrm{mag} \approx 10^{37}\,\mathrm{erg}\, \left(\frac{\epsilon}{0.1}\right) \left(\frac{B}{10^4\,\mathrm{G}}\right)^2 \left(\frac{\mathcal{F}}{0.3}\right)^{3/2},
\end{equation}
where $\epsilon$ is the fraction of magnetic energy $E_\mathrm{mag}$ convertible into flare energy, $B$ is the magnetic field strength of the sunspot, and $\mathcal{F} = A_\mathrm{spot}/\left(2\pi R_\odot^2\right)$ is the fraction of the Sun's surface covered by the sunspot with $A_\mathrm{spot}$ denoting its area; the normalization factor of $0.3$ was selected based on the preceding facts. Here, we have chosen a normalization of $10$ kG for the magnetic field as opposed to the standard value of $1$ kG \citep{Mae15}. Our choice is motivated by the fact that sunspots with $6$ kG are currently documented \citep{LHMW06}, while values $\lesssim 30$ kG have been conjectured for solar-type starspots \citep{RS00}. We have also normalized the efficiency by its characteristic value of $\sim 0.1$ \citep{KS13}.

Instead, if we suppose that $\alpha \sim 2$, the value of $E$ corresponding to $\tau$ is correspondingly increased by about an order of magnitude, i.e. it must have a value of $10^{38}$ erg. A superflare with this energy would not be feasible on the Sun, as the requirements for $B$, $\epsilon$ and $\mathcal{F}$ become very stringent and unlikely. Thus, we propose that, for the above choice of the parameters, the energy of the superflare that occurs once every $26$ Myr ought to be $E \sim 10^{37}$ erg. We have also suggested that superflares of this magnitude can, under a rare set of circumstances, occur on the Sun. We can also invert this argument as follows: upon computing the maximum possible energy $E$ of a solar superflare, we find that its frequency of occurrence corresponds to $\sim 20$ Myr. This value is very close to the periodic extinction timescale of $26$ Myr proposed by some authors \citep{RS84}. 

At this stage, a few important points regarding solar superflares merit a mention: (i) they may occur at much longer intervals than $\sim 20$ Myr \citep{Gop17}, and (ii) they could be unevenly spaced. The former stems from the fact that the flare distribution might decline rapidly due to an exponential falloff at large values instead of the power law scaling (\ref{Freq}). Such behavior has been documented for flux ropes that arise during the reconnection process \citep{Jan17,LCB17}. We note that (ii) can be partly explained by invoking the fact that superflares in certain G-type stars are not strictly periodic since they have been documented to occur in `clusters' \citep{Mae12,Shi13}. Moreover, as stellar activity broadly declines with age \citep{MaHi08,Sod10}, it is reasonable to expect that the frequency of superflares and the maximum energy released will reduce over time, implying that large superflares would tend to become increasingly uncommon during later epochs.

Hence, the above facts collectively indicate that the likelihood of superflares on the Sun being rare and intermittent, as opposed to regular and periodic, is also quite high. In turn, any extinction events they potentially cause would also display the same properties. Consequently, the hypothesis that some of the extinction events recorded since the Cambrian period \citep{HW97} could have been triggered by a superflare merits further consideration. Before proceeding further, we also wish to reiterate that the timescales discussed herein are subject to a fair degree of uncertainty as the statistical properties are not robust; instead, there is a paucity of data with respect to both the fossil record and superflares on solar-type stars. Hence, these timescales should be interpreted as the characteristic values associated with the corresponding processes.

Lastly, we observe that comparatively smaller superflares, i.e. with energies much lower than $10^{37}$ erg, may also play a role in regulating the biodiversity on Earth. Although these events are not expected to cause mass extinctions, their relative frequency is much higher compared to larger superflares \citep{Mae12,KS13}. Hence, it should be instructive to compare fossil biodiversity records against the predicted frequencies of superflares (with varying energies), and determine whether any significant correlations can be deduced.

\subsection{Caveats and assumptions for the model}\label{SSecCav}
Here, we shall elucidate the assumptions and uncertainties associated with our preceding discussion. 

We begin with the important observation that the maximum energy as well as the constant of proportionality and the spectral index of solar superflares in (\ref{Freq}) remains unknown. Hence, there is an inherent degeneracy that can be illustrated by the following example. Let us suppose that the timescale for a solar superflare with $10^{34}$ erg is $6.5 \times 10^4$ yrs, which is higher than our previous choice by a factor of $30$. Using $\alpha = 2.3$, we find that a superflare with $E \sim 10^{36}$ erg has a characteristic timescale of $\tau = 26$ Myr. In Sec. \ref{SSecOzDep}, we argue that even a superflare with $\sim 10^{36}$ erg has the potential to cause mass extinctions. A lower value of $E$ would, in turn, entail less stringent constraints on $\mathcal{F}$ and $B$ in (\ref{FlareEn}). 

Next, it must be recognized that our analysis is based on \emph{statistical} considerations. Hence, in employing (\ref{Freq}) we are implicitly relying on the assumption that the sampling of a large number of G-type stars is roughly equivalent to sampling the Sun over an extended period of time (around $4\times 10^5$ yrs). However, it must be noted that not all G-type stars are ``Sun-like''. This intrinsic variability may imply that the corresponding statistics for the Sun are \emph{not} the same as (\ref{Freq}). Hence, there is a distinct possibility that the Sun is incapable of giving rise to large superflares \citep{Sch12,KO16}. On the other hand, the analysis of young solar-type stars based on \emph{Kepler} data \citep{Shi13} suggests that the Sun would have been more active when it was much younger.

The estimate for the flare energy in (\ref{FlareEn}) constitutes a simple scaling analysis, and does not capture the time dependence of the flare energy released during the reconnection process \citep{ShiMa11,Pri14}. Moreover, it is important to recognize that the magnetic energy $\left(\propto B^2\right)$ is not fully converted into the flare energy since the final magnetic field is not completely available for explosive energy release; this necessitated the inclusion of an efficiency factor $\epsilon$ in (\ref{FlareEn}).

Lastly, we point out that the large value of the magnetic filling fraction $\left(\mathcal{F} \sim 10\%\right)$ employed in (\ref{FlareEn}) is necessary for large superflares to occur \citep{KS13}. The statistical analysis of the magnetic flux distribution on the Sun's surface indicates that such large values of $\mathcal{F}$ are not feasible \citep{MJ15}. An important point that needs to be noted, however, is that detailed solar observations of sunspots date back to a few centuries \citep{Sch12}, which is clearly a very short time span by geological and astronomical standards; in particular, the plasma environment of Earth during the Hadean and Archean epochs could have been quite different \citep{Aira16}.

\section{Effects caused by the superflare on Earth} \label{SecEff}
Next, we shall explore the environmental and biological consequences arising from a superflare with $E \sim 10^{37}$ erg, and whether these effects are severe enough to trigger a mass extinction.

We begin by estimating the energy $E_\oplus$ that is deposited by the superflare on Earth. If one assumes that the flare energy is emitted isotropically, we find
\begin{equation} \label{EnEarth}
    E_\oplus = E \left(\frac{R_\oplus}{2a}\right)^2,
\end{equation}
where $a = 1$ AU. In contrast, it has been suggested that the energy could be deposited in a non-isotropic manner with an opening angle of $24^\circ$ \citep{MT12,NH14}. If we consider this scenario, the energy deposited will be ${E}_\oplus' \sim 100 E_\oplus$. Upon substituting the appropriate values in (\ref{EnEarth}), we find that $E_\oplus \sim 4.5 \times 10^{27}$ erg and $E_\oplus' \sim 4.5 \times 10^{29}$ erg. 

\subsection{Ozone depletion and its consequences} \label{SSecOzDep}
The role of ionizing radiation, produced by flares and other catastrophic phenomena, on atmospheric chemistry and surface biology has been investigated quite extensively \citep{Dart11,MT11,AM14}. A number of factors, such as the Earth's thick atmosphere, the presence of a magnetic field, and the existence of ozone, serve to shield the surface from the majority of biologically damaging radiation - mostly Ultraviolet-B (UVB) and Ultraviolet-C (UVC). However, many of these studies concentrated on flares that were typically $< 10^{33}$ erg.  To the best of our knowledge, the effects of a flare with $E \sim 10^{37}$ erg do not appear to have been delineated in the literature. Hence, our subsequent discussion will necessitate a certain degree of extrapolation from known results. 

It is important to recognize that, in discussing the dangers arising from solar (or stellar) flares, there are several distinct components associated with the latter phenomena \citep{Schw06,ShiMa11,Em12,Benz17}; for e.g., the electromagnetic radiation emitted, and the high-fluence outflow of solar energetic particles (SEPs). The second factor has been explored in detail \citep{Mir01} and identified as being particularly important, since it facilitates the formation of nitrogen oxides $\left(\mathrm{NO}_\mathrm{x}\right)$ by means of atmospheric ionization \citep{Cru79}. In turn, these compounds are responsible for the depletion of ozone \citep{SH07,Jack08}. As noted earlier, reduction in the ozone levels enables higher doses of harmful UV radiation to reach the surface and leads to biological damage. We will now turn our attention to a few specific analyses of solar and stellar flares and their consequences.

The famous 1859 Carrington flare \citep{Carr59,CS04} remains one of the most powerful solar storms ever documented, with a total energy that is estimated to have been $5 \times 10^{32}$ erg \citep{CD13}. This event led to a globally averaged maximal decrease in the ozone levels of $5\%$, although maximum depletion was $14\%$ at higher latitudes \citep{TJM07}. Other studies found that ozone reduction of $20\%$-$40\%$ and $60\%$ occurred in the stratosphere and mesosphere respectively \citep{RVC08,Rod08,CUR13}; for comparison, the ozone depletion in the stratosphere because of anthropogenic change is $\lesssim 10\%$ \citep{Sol99}. We further point out that similar values for ozone reduction in the stratosphere and mesosphere were recorded for the 1972 \citep{HKC77}, 1989 \citep{JFV00}, 2000-2003 \citep{LF05,Jack05} and 2005 \citep{SV06} SEP events. In addition, significant changes in the surface air temperature were identified for the Carrington flare, with Europe and Russia experiencing warming of $\lesssim 7$ $^\circ$C \citep{CUR13}; the 2003 SEP event was responsible for temperature variations of up to $\pm 3$ $^\circ$C \citep{JR07}.

The terrestrial effects of the putative superflare in AD 775 were explored in \citet{TMAS}, and it was concluded that an SEP fluence of $\sim 10^{12}\,\,\mathrm{protons}\,\mathrm{cm}^{-2}$ of particles with energies $> 30$ MeV would lead to severe damage of the biosphere.  However, owing to the paucity of available data, a wide range of outcomes were predicted. The averaged ozone depletion ranged from a lower bound of $5\%$ to an upper bound of $32\%$ depending on the SEP fluence, with a fairly plausible intermediate value of $22\%$. In comparison, ozone depletion due to a GRB at a distance of a few kpc is $38\%$ \citep{TMJ05}, and a supernova at $8$ pc leads to a depletion of $47\%$ \citep{GLJ03}. The biotic effects due to the intermediate and upper cases were manifested as the increase in UVB-induced damage of plants by $14\%$ and $25\%$ respectively. The SEP event in AD 775 was therefore associated with moderate damage of the biosphere due to reduced photosynthesis in the oceans and land \citep{TMAS}.

We also note that the effects arising from strong flares have been studied for Earth-analogs orbiting M-dwarfs. \citet{SWMKH} considered a superflare on the active M-dwarf AD Leonis (AD Leo), and demonstrated that the UV radiation did not cause any significant ozone depletion. However, when the role of SEPs was taken into account, the ozone depletion was shown to attain a maximum of $94\%$. Here, two caveats must be recorded: the Earth-analog was situated at a distance of $0.16$ AU and the maximum value was for an unmagnetized planet. 

Given that only a few data points are available, all results obtained from direct extrapolation must be interpreted with due caution. Furthermore, there are several other factors involved in the extent of ozone depletion, for e.g. diurnal cycles \citep{Verr05}, which are not considered in our analysis. Let us denote the ozone depletion by $\mathcal{D}_{O_3}$ and the SEP fluence by $\mathcal{F}_p$. Assuming a power-law scaling, we suggest that the following expression serves as a reasonable fit for the SEP events discussed earlier:
\begin{equation} \label{OzDep}
   \mathcal{D}_{O_3} \sim  5\%\,\left(\frac{\mathcal{F}_p}{10^{10}\,\,\mathrm{protons}\,\mathrm{cm}^{-2}}\right)^{2/5}.
\end{equation}
The next step is to relate the flare energy $E$ to the SEP fluence. To do this, we invoke the results from \citet{TMS16}, where the scaling of the SEP \emph{flux} $F_p$ with $E$ was obtained. Assuming an isotropic angular distribution of the SEPs, and using $F_p \propto \left(V_{CME}\right)^{4.35}$, $V_{CME} \propto E^{1/6}$, $t_{CME} \propto E^{1/6}$ and $\mathcal{F}_p \propto t_{CME} \times F_p$, the scaling relation is
\begin{equation} \label{FluEn}
    \left(\frac{\mathcal{F}_p}{10^{10}\,\,\mathrm{protons}\,\mathrm{cm}^{-2}}\right) = \left(\frac{E}{5 \times 10^{32}\,\mathrm{erg}}\right)^{9/10},
\end{equation}
where we have normalized the fluence and flare energy in terms of the Carrington event \citep{CD13}. Upon combining (\ref{OzDep}) and (\ref{FluEn}), we arrive at
\begin{equation} \label{OzEn}
   \mathcal{D}_{O_3} \sim  5\%\,\left(\frac{E}{5 \times 10^{32}\,\mathrm{erg}}\right)^{9/25}.
\end{equation}
If we substitute $E \sim 10^{37}$ erg in the above expression, we find that $\mathcal{D}_{O_3} \sim 177\%$. As noted earlier, the same power-law behavior may not be valid for higher flare energies and SEP fluences. It is also possible to compute the critical energy $E_c$ that leads to $100\%$ ozone depletion; we find it to be $E_c \sim 2 \times 10^{36}$ erg. If the flare on AD Leo were scaled upwards to account for the larger Earth-Sun distance, its equivalent energy would be $\sim 10^{35}-10^{36}$ erg. As this flare caused a maximum of $94\%$ ozone reduction \citep{SWMKH}, the flare energy is roughly in agreement with the value of $E_c$ calculated from our model. 

Let us recall that a GRB from a few kpc leads to ozone depletion of $\sim 40\%$ \citep{TMJ05} and has been posited as the trigger for the Ordovician mass extinction \citep{Mel04}. A supernova at $8$ pc has also been predicted to engender comparable depletion \citep{CB96,GLJ03}. In contrast, as per our scaling relations, a flare energy upwards of $E_c$ would cause complete destruction of the ozone layer and correspond to a fluence of $\sim 10^{13}\,\,\mathrm{protons}\,\mathrm{cm}^{-2}$. If such a superflare were to occur on the Sun (regardless of its periodicity), it seems reasonable enough to argue that the damage to the biosphere would be great enough to trigger a mass extinction, especially since severe ozone depletion engenders widespread and major biological damage \citep{TNS15}. Hence, if a large flare (even one with $E \ll E_c$) subsequently erupted before the ozone layer had been replenished, virtually all organisms on the surface, including extremophiles, would be critically endangered \citep{EV17}.

We will now briefly summarize some of the effects that arise due to ozone depletion. Ozone depletion has been linked with the increased penetration of biologically harmful UVB radiation \citep{KM93}, an environmental stressor that leads to mutagenesis, reduced fertility, suppression of physiological processes, and even death \citep{VR93,DJ10}. Recent research suggests that the primary influences of UVB radiation on life are likely to be manifested at trophic levels - moving the focus away from individual organisms and species - through alterations of biogeochemical and climate cycles \citep{CWLA,Had07,Zepp11}. For instance, UVB radiation may indirectly cause a reduction in carbon dioxide absorption, or a decline in the quantity/quality of nutrient cycling in marine food webs \citep{Had15}.

As a specific example, we point out that enhancement of UV radiation leads to a reduction in phytoplankton photosynthesis \citep{CN94,DN02} and causes DNA damage \citep{Mal97,CGP12}. The ozone depletion in the Antarctic has been linked with a $\gtrsim 10\%$ decline in the productivity of phytoplankton \citep{SP92}. Any such decline would have crucial effects on marine ecosystems since phytoplankton are responsible for $50\%$ of the planet's primary production \citep{FB98}. In addition, phytoplankton play a critical role in regulating biogeochemical cycles, climate variations, biomass production, and the diversity, abundance and functioning of marine ecosystems \citep{CWLA,Sab04,BLW}. 

Any changes in plankton productivity will cause ripple effects that extend to different trophic levels, and thus alter the overall ecosystem response to UVB radiation \citep{BSP94} as discussed in the preceding paragraph. A decline in phytoplankton could, in principle, disrupt the biological pump and lead to increased CO$_2$ levels in the atmosphere and an associated rise in temperature via the greenhouse effect. We observe that some, albeit not all, of these factors have been documented for the Permian-Triassic mass extinction \citep{KBCG96,Erwin06,KBPPF} although this does \emph{not} necessarily imply that the P-T extinction was triggered (or exacerbated) by a superflare.

\subsection{Other ramifications from the superflare}\label{SSecOth}
Apart from the manifold consequences of sudden ozone depletion and enhanced UVB radiation, superflares of this magnitude could also give rise to other effects, some of which have an interesting mix of negative and positive consequences. As described earlier, \citet{CUR13} concluded that the the Carrington event raised surface air temperatures by $7$ $^\circ$C. Here, it is worth recalling that the superflares we consider are approximately five orders of magnitude larger. Hence, it seems reasonable to conclude that the air temperature would be subjected to a much higher increase (or decrease). Although this rise (or fall) in temperature would be transient, we hypothesize that this could have a highly detrimental effect on most complex organisms for several reasons.

First, we observe that most organisms have an optimal body temperature at which they function. If the temperature exceeds this value by a non-trivial amount, there is a sharp decline in biochemical and physiological processes, ultimately leading to protein denaturation \citep{Sch15}. Another factor that is even more important than the rise in temperature is the \emph{timescale} over which it occurs. If the spike in temperature is sharp, the organism's thermal adaption breaks down \citep{Ang09}. Hence, it appears reasonable to conclude that the metabolic functioning of most organisms would be impaired, perhaps irreversibly, when subjected to a superflare. As temperature regulates a wide array of ecological and evolutionary properties \citep{BGASW,LiLo17}, we anticipate that an abrupt increase in temperature would severely impact the stability and functioning of ecosystems. Rapid fluctuations along these lines have been posited as major causes behind the ongoing Holocene extinction \citep{Barn12}. It is therefore reasonable to surmise that past mass extinctions could also have featured elevated temperatures \citep{KS05}.

Nitric acid rain is generated through the reaction of nitrogen dioxide $\left(\mathrm{NO}_2\right)$ with the hydroxyl group \citep{Cru79,Toon86}. The ensuing nitrogen pollution of aquatic ecosystems leads to a multitude of issues including acidification, increased toxicity, and eutrophication \citep{CA06}. On the other hand, since it can lead to a proliferation of primary producers, it may enable ecosystems to rebound after the initial destructive phase. Similarly, it has been argued that $\mathrm{NO}_2$ can reduce solar irradiance and cause large-scale glaciation \citep{RMC78}. However, recent studies indicate that this phase is transient and accompanied by a subsequent increase in solar irradiance \citep{TNS15}. As some of the mass extinction events appear to have been followed by an upsurge in species diversification \citep{Ben09,Knoll15}, factors with dual characteristics, like the ones identified above, might have played an important role. 

Before proceeding further, we wish to highlight a couple of self-evident, but nonetheless highly important, conceptual points: ecosystems, as well as the biosphere, are intrinsically nonlinear. The study of nonlinear dynamical systems has revealed the significance of ``tipping points'', i.e. states wherein infinitesimal perturbations can give rise to critical transitions leading to qualitative changes \citep{Len08}. Hence, even when considering cases where superflares give rise to only ``minimal'' changes in the environment, Earth's climate and biosphere may respond in a nonlinear manner \citep{SCFFW}, thereby possibly leading to the onset of a mass extinction event.

In the same spirit, we advocate that astrophysical causes should not be viewed in isolation, as they are more effective when acting in tandem with geological phenomena, for e.g. geomagnetic field reversals, volcanism and ocean circulation patterns. The Earth's magnetic field is significantly reduced during the reversal process \citep{MMM}, and an SEP event of lower magnitude occurring during this period will therefore be capable of causing the same degree of devastation \citep{RI76,Raup85}. Thus, superflares may constitute one half (the impulse) of the proposed ``press-pulse'' mechanism for mass extinctions  \citep{AW08}.\footnote{In many cases, however, the distinction between ``press'' and ``pulse'' is not readily apparent, and both give rise to a wide range of macroevolutionary responses \citep{GGH17}.} In turn, this could lead to mass extinction events that display a superposition of stochasticity and periodicity; such patterns have been predicted to be duly manifested in the fossil record \citep{Feu11}.

\subsection{Signatures of solar superflares}
Having outlined the consequences arising from a massive superflare, it is now instructive to ask whether such flares can be deduced from the geological record. 

As noted previously, one of the consequences of superflares is that they can give rise to large-scale SEP events that promote the production of nitrogen oxide compounds. The most widely proposed method entails the use of ice cores in Greenland or Antarctica as a proxy for solar activity \citep{Sto80}. The basic idea is that there exists a correspondence between nitrate concentrations in the ice cores and flares; short-term nitrate features (spikes) can, in principle, reflect solar proton events \citep{LK90,DZ98,McD01}. However, when considering studies reliant on this method, it is important to distinguish between natural (for e.g. flares) and anthropogenic deposition of nitrates, since the latter has become increasingly important \citep{May86}.

There has been some controversy as to whether SEP events are large enough to account for the observed peaks. For instance, even the highly energetic Carrington event has not left widespread traces in polar ice \citep{Wolff12}. Several authors have therefore concluded that solar proton events in the Holocene $\left(\sim 10^4\,\mathrm{yrs}\right)$ are not detectable, implying that nitrate spikes are not accurate proxies for these phenomena \citep{Wolff12,Dud16}. In contrast, it has recently been suggested that hard-spectrum SEP events can be unambiguously identified \citep{Smart14,Mel16} in the ice core record. As the superflares considered in our work are much stronger than those recorded in modern history, the level of nitrate deposition due to the associated solar proton events should be much higher; on the other hand, their age ($\sim 10$ Myr) and the accompanying erosion may render this method non-viable. 

From a long-term standpoint, it seems probable that the use of isotope-based estimates represents a more promising endeavour. In particular, it has been suggested that SEPs can produce $^{10}$Be and $^{14}$C in the atmosphere, and that high-resolution isotope data may yield signatures of such events \citep{Uso06,Beer13}. A rapid increase in the $^{14}$C content of tree rings in Japan \citep{MNMN} has been invoked as evidence in favour of a high-fluence solar proton event in AD 775 \citep{MT12,Uso13,Mek15}. Thus, if abrupt features are present in measurements of cosmogenic radionuclides at the same period as one of the mass extinction events, they would lend credence to the hypothesis that superflares played a role in triggering species extinction; note that these spikes must also be consistent with a solar proton event. In order to evaluate this conjecture, high-resolution data pertaining to these radionuclides should be collected from both terrestrial and lunar rocks.

We conclude by observing that evidence from radionuclides and ice cores in favor of superflares must be interpreted with due caution. One must identify potential ``false positives'' that are capable of producing the same signatures as solar superflares, and may therefore be mistaken for the latter. It is safe, however, to argue that the aforementioned radiochemical evidence does not unequivocally eliminate the possibility of solar flares.\footnote{The identification of false positives also constitutes an important component in the analogous field of detecting biosignatures on exoplanets \citep{Kal17}.}

\subsection{Imprints in the fossil diversity record} \label{SSecImp}
We begin by observing that the putative causes for the `Big Five' mass extinctions have been quite thoroughly documented \citep{HW97,Cour99,Bam06}. Hence, the chances of superflares causing these particular extinction events are most likely minimal. However, as noted in the previous sections, some of the observed extinction events with a periodicity of $26$ Myr may have been caused by superflares, often acting in conjunction with other natural causes. We will therefore outline certain distinctive features that might characterize extinctions where superflares played a role.
\begin{enumerate}
    \item Ozone depletion is predicted to increase (up to a factor of $2$-$3$) as one moves to higher latitudes \citep{TJM07}. In light of the deleterious consequences of ozone depletion outlined in Sec. \ref{SSecOzDep}, we may expect the extinction probability to increase with latitude. We also note that the temperature spike described in Sec. \ref{SSecOth} is likely to be more pronounced at higher latitudes. In contrast, the opposite (extinction probability) trend has been predicted to occur for extinction driven by current climate change \citep{TCG04}. 
    \item Organisms that are subterranean or those that dwell below the euphotic zone should be relatively protected from UVB radiation. Hence, the preferential extinction of surface-dwelling (land or aquatic) organisms could be a consequence of superflares, unless they are equipped with screening compounds \citep{CK99}.
    \item The photosynthetic productivity of phytoplankton has been argued to be more sensitive to UVB levels when compared to terrestrial plants due to its less effective screening \citep{DN02}. Given the importance of the former in oceanic ecosystems \citep{FB98}, we conjecture that aquatic species would be rendered more vulnerable to extinction when compared to terrestrial organisms.
    \item Nitric acid rain and the production of nitrogen oxides are some of the outcomes that may result from superflares, as discussed in Sec. \ref{SSecOth}. Their ensuing consequences have already been investigated in the context of the P-Tr and K-Pg extinctions \citep{PF87,Zah90}, and include photosynthesis inhibition, toxicosis, foliage and respiratory damage. Invertebrates in freshwater ecosystems are particularly vulnerable to acidification \citep{Sch88}, and might therefore be preferentially subject to extinction compared to their saltwater counterparts.
\end{enumerate}
The above list is not meant to be definitive, but it can potentially serve as a preliminary guide for locating extinction events mediated by superflares.

\section{Implications of superflares for life elsewhere in the Universe}\label{SecImp}
We will now explore some of the implications that superflares would have for life-as-we-know-it on other planets (and moons).

\subsection{Implications for Mars and Venus}
Here, we shall consider only present-day Mars and Venus, and return to ancient Mars and Venus at a later stage. If seen purely from the viewpoint of energetics, it may appear as though the UV and particle energies deposited on Mars (Venus) are only a factor of $2$ lower (higher) than those deposited on Earth, provided that the emission is isotropic with an inverse-square law.  

In reality, the scenario is more complicated on account of the fact that Mars has a very tenuous atmosphere - the surface pressure and column density are about two orders of magnitude lower \citep{Owen92} - and weak (crustal) magnetic fields \citep{Acu98}. Both of these factors have been identified as major obstacles in protecting the surface from Galactic Cosmic Rays (GCRs). For instance, while considering exoplanets around low-mass stars, \citet{GTV15} concluded that the presence of a weak (or zero) magnetic field leads to an energetic particle flux that is more than three orders of magnitude higher than on Earth. Hence, strong magnetospheric shielding is necessary, especially for planets with rarefied atmospheres, to prevent elevated surface radiation levels \citep{GTS16}.

As Mars possesses these characteristics, it seems reasonable to conclude extremely large superflares of the kind discussed herein would prove to be highly detrimental, and possibly fatal, to any life on the planet. Several studies have attempted to trace the evolution of Martian habitability over time \citep{CCD00,Fair10,Cock14}, and identify regions where life could have persisted \citep{Bos92,DSM16}. After the discovery of superflares in G-type stars \citep{Mae12}, detailed studies of Martian habitability when subjected to such an event do not seem to have been undertaken. Although the radiation doses are unlikely to drive \emph{all} Martian lifeforms to extinction, any survivors would have evolved a high radiation tolerance, akin to organisms like \emph{Thermococcus gammatolerans} \citep{Jol03}, \emph{Deinococcus radiodurans} \citep{CB05} and \emph{Milnesium tardigradum} \citep{Hor06}.

The situation for Venus is quite different. In this case, studying the role of surface ionizing radiation is quite irrelevant since the temperature (at $740$ K) is far too extreme to host life-as-we-know-it. Instead, proposals for putative Venusian life have focused on sulphur-based chemoautotrophs situated in the clouds \citep{Mor67,Cock99,SM04}. The effects of the AD 775 SEP event on the Venusian atmosphere were studied by \citet{Dart15}, and it was concluded that the radiation dose was insufficient to cause damage, although the atmospheric chemistry was affected by the strong ionization. By using the scaling relation (\ref{FluEn}), we conclude that SEP fluence would be nearly four orders of magnitude higher for the superflares discussed in this paper when compared to the AD 775 event. Hence, it seems likely that airborne ecosystems on Venus would be subject to high extinction risks when these superflares occur due to a much higher degree of ionization and greatly enhanced radiation dosage.

In addition, a wide range of objects in our Solar system have been proposed as sites where life could exist \citep{SI06,SSM09}; the list includes well-known candidates like Titan, Europa and Enceladus, but also more exotic options such as Jupiter \citep{SS76}, asteroids and comets \citep{CB99}. For subsurface environments, we do not anticipate that superflares would play a major role provided that the crust is sufficiently thick; see, however, \citet{Dart11}. Atmospheric ecosystems, on the other hand, are likely to be significantly perturbed by major superflares along the lines described earlier.

\subsection{Implications for life on exoplanets around M- and K-dwarfs}\label{SSecMD}
We begin by observing that planets situated in the habitable zone (HZ), the region theoretically capable of supporting liquid water, around M-dwarfs are characterized by two distinct and highly important properties: (i) they are situated very close to the host star, and (ii) the host stars are very active \citep{SK07}.

A combination of these two factors is responsible for ensuring that the atmospheres of M-dwarfs are rapidly stripped away through a combination of thermal and non-thermal escape processes \citep{DLMC,LiLo17,GS17}. Simulations have illustrated that these mechanisms lead to atmospheric depletion over $< 1$ Gyr timescales in the absence of outgassing. Most of the exoplanets in the HZ of M-dwarfs are tidally locked, and have weak magnetic moments \citep{Kho07,Zul13}. On account of these two reasons, the shielding against coronal mass ejections, SEPs and GCRs is expected to be much lower \citep{Grie05,Grie09,Vid13,GTV15,Kay16}. Consequently, the biological hazards will be heightened for planets in the HZ of M-dwarfs. For exoplanets orbiting active M-dwarfs, it has been shown recently that the levels of surface UVB and UVC radiation (due to flares) would be lethal to most lifeforms on Earth \citep{OMJK17}.

\citet{Atri17} studied the effects of solar proton events on a wide range of Earth-analogs with varying orbital distances, magnetic moments, atmospheric column densities and flare energies. For the most extreme case(s), the radiation doses on the surface were demonstrated to be $\sim 10^4$ Sv. In comparison, a dose of $\sim 100$ Sv is lethal to most mammals and birds, and certain insects. Most studies concerning the biological ramifications of superflares have hitherto assumed a maximum flare energy of $\lesssim 10^{36}$ erg. However, as per our discussion in this paper, the existence of more energetic superflares ought not be ruled out. M-dwarfs have a smaller surface area compared to G-type stars, and thus smaller (maximal) spot sizes, but they also have higher surface magnetic fields \citep{Morin10}, and may therefore still be capable of generating large superflares.

Apart from the greatly enhanced radiation doses received on these planets, we note that superflares with $\sim 10^{35}$ erg have an occurrence rate that is $20$ and $5$ times higher for M- and K-dwarfs respectively compared to G-type stars \citep{Mae12,Cand14}. Hence, complex life on exoplanets around M-dwarfs could be subject to repeated extinction events on the timescale of $\mathcal{O}(1)$ Myr. In between these events, we observe that smaller superflares occur at regular intervals, potentially lowering the chances for the biosphere to repair itself. However, in light of the enhanced mutations and selection pressure induced by flares \citep{Sag61,Sag73,Cock98,Dart11}, the periods in between these extinctions may witness rapid speciation. Superflares could therefore be responsible for periodically varying diversification and extinction rates. Thus, it seems plausible that short bursts of extinction and speciation \citep{SSW04} might be interspersed with long periods of stasis; the suggested pattern is somewhat reminiscent of punctuated equilibrium \citep{EG72,GE93}. 

Let us now suppose that we consider the idealized scenario where all of the energy from the superflare impacts the surface of an Earth-sized planet orbiting a low-mass M-dwarf. The energy deposited $E_p$ is estimated by utilizing (\ref{EnEarth}) and we will choose $a \sim 0.01$ AU for the sake of convenience; this value is somewhat close to the orbital radii of Proxima b \citep{AE16} and the TRAPPIST-1 planets \citep{Gill17}. We find that $E_p \sim 4.5 \times 10^{31}$ erg for isotropic emission, and $E_p \sim 4.5 \times 10^{33}$ erg for non-isotropic emission with an opening angle of $24^\circ$. We ask the question: what is the mass $M$ that will be raised to the boiling point of water? It is computed via
\begin{equation}
    M = \frac{E_p}{\mathcal{C} \Delta T},
\end{equation}
where $\Delta T \sim 100$ K and $\mathcal{C}$ is the specific heat capacity of water. In reality, note that all of the energy impacting the planet will not be delivered to the surface, and the value of $\Delta T < 100$ K. With these values, we find that $M \sim 10^{19}$ kg for isotropic emission, and $M \sim 10^{21}$ kg for the non-isotropic case. This leads us to the remarkable conclusion that, for the latter situation, a superflare of $10^{37}$ erg is capable of evaporating the oceans on this planet provided that their total mass is comparable to that of Earth's oceans. Thus, in terms of an existential threat, it should be placed in the same category as asteroids, GRBs and supernovae \citep{SBL17} although its frequency of occurrence is much higher. Even though a single superflare will not suffice to wholly evaporate the oceans, a few of them, spanning a total of $\mathcal{O}(100)$ Myr, should be enough to dessicate a planet in the HZ of a low-mass M-dwarf.

If we consider the isotropic emission case, the ramifications are still severe, albeit not so dramatic. The euphotic zone, the region where photosynthesis occurs and most of the marine life is situated, would be completely evaporated, and the same fate would befall the rest of the pelagic zone. Even on Earth, the non-isotropic scenario is capable of raising the temperature of the photic zone by a few degrees and could disrupt biogeochemical mechanisms and giving rise to outcomes like euxinia; the latter is believed to have played an important role in regulating ocean diversity over time \citep{MK08}, and in the Permian-Triassic mass extinction event \citep{Gri05}.

Thus, to summarize, the prospects for complex life on exoplanets in the HZ around M-dwarfs are severely hampered due to a multitude of reasons. The degree of ozone depletion and the radiation dosage received are likely to be much higher than those on Earth. Superflares occur with a higher frequency on M-dwarfs and are thus more likely to give rise to frequent extinction events. Lastly, they could deposit enough energy into the oceans to boil them completely or partially, and thereby severely impact the growth and development of marine life. Although our discussion was oriented towards exoplanets, many of these considerations would be applicable to exomoons in the HZ \citep{HW14} as well. Some of the general conclusions regarding planets orbiting M-dwarfs are applicable to K-dwarfs to a lesser degree, as the latter fall between M- and G-type stars in terms of most of their properties.\footnote{Since superflares are known to exist even on L-dwarfs \citep{Schmidt}, we anticipate that our findings would also be valid to some degree to planets orbiting such stars.}

Collectively, these facts pave the way towards answering a fundamental question delineated in \citet{LBS16}: why is it that we orbit a G-type star in the present epoch and not an M-dwarf in the cosmic future? This question was further studied through the use of Bayesian inference methods \citep{HKW17}. One approach to resolve this apparent paradox is by identifying reasons why life around M-dwarfs is selectively suppressed. Through considerations of biodiversity, \citet{Ling17} recently argued that low-mass M-dwarfs are unsuitable for life-as-we-know-it, implying that K- and G-type stars represent the best chances for hosting complex biospheres \citep{HA14,CG16}. Although this conclusion ameliorates the problem, it does not fully solve it since we are left with the equivalent question: why do we orbit a G-type star and not a K-dwarf? We suggest that superflares might represent a missing piece of the puzzle: their impact on exoplanets in the HZ of K-dwarfs is more profound, and these events occur more frequently (by a factor of $5$). Thus, when all of these factors are taken into consideration, our position around a G-type star may not be a fortuitous accident, but a fairly probable event instead. 

If a large fraction of M- and K-dwarfs are unsuited to host complex life on planets orbiting them, this still leaves G-type stars. However, even in this category, we note that a small, but non-trivial, fraction of them display evidence of regular superflare activity \citep{Mae12,Shi13}. One may thus be tempted to conclude that complex life is rare in the Universe, although simple microbial life could be quite common. This line of reasoning has been advocated by several authors in the past and is referred to as the ``Rare Earth'' hypothesis \citep{WaB00}. However, we wish to caution that our study does not necessarily direct us to this conclusion since the fraction of G-type stars exhibiting unusual superflare activity is known to be small, and there exist considerable statistical uncertainties regarding the frequency and magnitude of superflare events on M-, K- and G-type stars.

\subsection{Risks to human civilization from superflares}
Ever since the discovery of superflares in G-type stars, several studies have briefly alluded to the risk to human civilization from such an event \citep{KS13,KK16}. However, detailed analyses of the threats posed by a large superflare to \emph{technological} civilizations (such as ours) do not appear to have been undertaken thus far \citep{LiMa17}.

In Sec. \ref{SecSync}, we presented data favoring the recurrence of a $\sim 10^{34}$ erg superflare every $\sim 2000$ yrs. Moreover, superflares with energies of approximately $10^{35}$ erg, $10^{36}$ erg and $10^{37}$ erg would occur with frequencies of $\sim 40$ Kyr, $\sim 800$ Kyr and $\sim 20$ Myr respectively. Even though superflares with relatively lower energies will cause negligible biological damage, they \emph{are} capable of causing tremendous destruction to human civilization. Hence, it is imperative to constrain (and eventually predict) the frequencies with which these superflares can occur on the Sun. The first step entails undertaking a thorough scrutiny of historical records for evidence of large-scale aurorae and sunspots that could be indirectly associated with superflares. Although some studies along these lines have been undertaken recently \citep{Vaq07,Hay15,Hay17,Tam17}, a much higher degree of attention to this topic appears to be warranted.

We also note that several studies have attempted to forecast the course of space weather over the next few centuries \citep{Barn11,Lock12,SB13,IM15} but most of them have focused on making predictions over short timescales, i.e. for the next $10-100$ years. For instance, models indicate that a Carrington-like event has a relatively high ($\lesssim 10\%$) chance of occurring in the next decade \citep{Show11,Ril12,Kat13}. The emphasis on short timescales is motivated primarily by pragmatic considerations since the inherent solar variability does not enable accurate forecasting over longer epochs. However, as we shall argue below, there is a pressing need to take longer timescales into account.

We begin by observing that the manifold impacts of relatively moderate (in comparison to superflares) space weather events have been thoroughly documented \citep{Schw06,SSB09,Hap11,SKA15,East17}. Coronal mass ejections, typically associated with flares, give rise to powerful geomagnetic storms capable of significantly disrupting the planet's magnetosphere \citep{Kahl92,WH12}. Geomagnetic storms induce large electric fields and currents, which can severely disrupt a wide range of electrical systems \citep{BPN98,Pir00,Pul07}. A superflare may also generate an electromagnetic pulses (EMP) due to the abrupt ionization of the planet's dayside atmosphere, somewhat akin to the effects of a nuclear weapon \citep{GD77,Long78,Voll84}. Detailed calculations pertaining to these processes, and the ensuing consequences for technological civilizations, are beyond the scope of this paper.

The Carrington 1859 flare has garnered much attention since it represents a valuable benchmark against which extreme space weather events can be measured. In 1859, the Carrington flare caused the disruption of telegraph services \citep{Bot06}, but the same event would lead to far more destructive effects in the current era. For starters, we note that the worldwide disruption of power grids would lead to considerable economic damage. The losses for the US alone have been documented to be $\sim 2$ trillion dollars \citep{SSB09}. In addition, breakdowns in satellite communications, navigation and surveillance are anticipated. The total economic losses have been estimated to be $\sim 70$ billion dollars, and about $10\%$ of the existing satellites orbiting the Earth would be destroyed \citep{OGT06}. More devastatingly, \citet{Sch14} concluded that the long-term disruptions of global supply chains due to extreme space weather events would lead to losses worth $3.4$ trillion dollars. Assuming these estimates are correct, the resulting impact would be equivalent to the cumulative effects of anthropogenic climate change over a period of several decades.\footnote{\url{http://www.ghf-ge.org/human-impact-report.pdf}}

In addition, we note that the SEPs produced during extreme space weather events constitute a major hazard to any space-based operations. Hence, in the roadmap of \citet{SKA15}, the need for further observations and modelling of SEP events was identified as one of the highest priorities. We also note that solar proton events damage the atmosphere by inducing chemical changes, disrupting climate feedback mechanisms, triggering electrical discharges and altering the formation of clouds \citep{GB10,SKH13,Mir15}. Each of these environmental changes will, in turn, also lead to concomitant ecological, social and economic losses that are likely to be quite significant.

Although the scaling between economic losses and the magnitude of catastrophes will \emph{not} be linear, it is still instructive to evaluate the energy of a superflare that would lead to damage equal to that of the world's Gross Domestic Product (GDP). Using the values for the Carrington event described above, and the world's current GDP,\footnote{\url{http://data.worldbank.org/indicator/NY.GDP.MKTP.CD?year_high_desc=true}} we find that the resultant value is $\sim 10^{34}$ erg. A superflare with this energy could occur on the Sun once every $\sim 2000$ years. If we further assume that the AD 775 event was a superflare of this magnitude, we are led to the conclusion that the next such event might take place $\sim 750$ years in the future. However, as noted in Secs. \ref{SSecT} and \ref{SSecCav}, our understanding of solar superflares is both rudimentary and based on statistical evidence. Hence, the wait time of $\sim 750$ years proposed above must be viewed with caveats.

Based on the frequency of superflares outlined earlier, we surmise that a event with energy $\sim 10^{36}$ erg has a $\sim 10^{-4}$ chance of occurring in the next century. As noted in Sec. \ref{SSecOzDep}, a superflare of this magnitude may be sufficient to cause total ozone depletion and lead to major ecological damage. In comparison, the likelihood of a $2$ km asteroid or comet hitting the Earth in the same period has been estimated to be $10^{-4}$ \citep{CM94} and would result in widespread destruction \citep{Toon97}. Hence, both of these events represent genuine hazards, and have a similar likelihood of occurring in the next century. However, despite the similar (or greater) dangers posed by superflares, asteroid and comet impacts have been subjected to detailed risk analyses \citep{Pos04,Smith13}. NASA has also put together extensive plans entailing the close monitoring of near-Earth objects, and deflecting them if necessary; the Panoramic Survey Telescope and Rapid Response System (Pan-STARRS) mission merits a mention in this regard.\footnote{\url{https://panstarrs.stsci.edu/}}. The total cost of a system to detect and deflect near-Earth objects is between $1-10$ billion dollars.\footnote{\url{https://www.nasa.gov/pdf/171331main_NEO_report_march07.pdf}} Although this discrepancy is partly explained by the recent discovery of superflares in G-type stars, it can also be attributed to ``anthropic shadow'' or cognitive biases that lead to underestimation of risks posed by certain catastrophes \citep{Y08,CSB10}.

We end our analysis with a brief comment on the well-known ``Doomsday argument'' which relies on probabilistic considerations to arrive at the total number of human beings who will exist in the future. \citet{Gott93} undertook a famous analysis that led to an estimate of the future lifetime of humans; the value ranged from $\sim 5 \times 10^3$ years to $\sim 8 \times 10^6$ years. Although Gott's analysis has been critiqued by several authors \citep{Bos02}, we can use these numbers to estimate the corresponding magnitude of the solar superflares by using (\ref{Freq}). We find that the superflare energies must lie between $2 \times 10^{34}$ erg and $6 \times 10^{36}$ erg; based on the arguments provided in this paper, the latter value is capable of causing a mass-extinction event.

\subsection{Some positive implications of superflares}
Hitherto, we have restricted ourselves to exploring the negative consequences arising from superflares. However, as briefly noted in Sec. \ref{SSecMD}, these flares stimulate mutations and thereby lead to bursts of rapid species diversification - a factor that may have been particularly important during the Archean era.

Even if one supposes that solar superflares gave rise to mass extinctions, polyextremophiles like \emph{Deinococcus radiodurans} would be easily able to survive such episodes. Moreover, the remarkable discovery of \emph{Desulforudis audaxviator} \citep{Chiv08}, a sulfur-reducing chemoautotroph, has revealed that species on Earth and elsewhere can derive energy from radioactive sources for sustenance. Hence, even in high-radiation environments that would result from superflares, a fair number of species may possess UV radiation screening and DNA repair mechanisms \citep{CK99}, and prove to be adaptable enough to survive. In addition, certain habitats, associated with reduced levels of UV radiation, ought to be conducive to the sustained existence of photoautotrophs \citep{CM98,Cock04}.

Superflares could have played a beneficial and important role during the Hadean era through a number of channels. SEPs, as well as GCRs, have the capacity to catalyze cloud formation \citep{Kirk07}, the generation of strong electric fields, significant bursts of energetic radiation, and lightning \citep{Dwy03,DU14}; the last has important biological consequences \citep{EW10}, given its relevance in prebiotic chemistry \citep{CS92}. Furthermore, SEPs impacting the Earth during such events would have enabled a network of chemical reactions \citep{EIB02}, ultimately culminating in the formation of nitrous oxide and hydrogen cyanide \citep{Aira16}. The former's importance stems from the fact that it is a highly efficient greenhouse gas that could have warmed Earth's atmosphere, thereby providing a potential resolution for the long-standing faint young Sun paradox \citep{SM72}.

It is, however, the latter compound that has attracted a great deal of attention in recent times. In studies concerning the origin of life, the `RNA world' hypothesis has been extensively investigated \citep{Joy02,RBD14}. The formation of activated ribonucleotides, which undergo polymerization to yield RNA, is difficult for a number of reasons. However, it was shown by \citet{PGS09} that a mixture of chemical compounds, including cyanamide and cyanoacetylene, led to the synthesis of pyrimidine ribonucleotides under conditions resembling the early Earth. Subsequently, \citet{Pat15} demonstrated that the forerunners of the building blocks for protocells - nucleic acids, proteins, and lipids - could have arisen through the homologation of hydrogen cyanide and its derivatives. Hydrogen cyanide has therefore been identified as a putative `feedstock' molecule which played a pivotal role in the origin of life \citep{Sal12,Suth16}. 

We note that the synthesis of these compounds need not have occurred on Earth since early Mars \citep{Word16} and Venus \citep{Way16} were also potential sites of prebiotic synthesis \citep{CSC00}. Asteroids and comets could have facilitated the exogenous delivery of prebiotic compounds to Earth by means of quasi-panspermia \citep{CS92,THCM06}. Looking further afield, we anticipate that planets orbiting M-dwarfs, and K-dwarfs to a lesser extent, would be more conducive to exogenous delivery mechanisms primarily on account of the shorter interplanetary distances involved \citep{Linga17}.

Lastly, flares lead to elevated levels of UV radiation, and have thus been invoked as a means of ameliorating the UV deficiency \citep{Buc07,Rug15} on planets around M-dwarfs. It was noted in \citet{Ran17} that UV-sensitive prebiotic chemistry pathways could be functional over the duration of the flare, and become inactive during the quiescent phase. These findings may be valid to some degree for superflares, although they have a much lower frequency of occurrence. Additionally, it seems quite plausible that these extreme phenomena can adversely impact the synthesis of prebiotic compounds once a certain threshold value of the energy (and UV flux) is exceeded.

\section{Conclusions} \label{SecConc}
Ever since the discovery of superflares on solar-type stars, there has been much interest in exploring the ensuing consequences of such events on Earth and other exoplanets. We began our analysis by proposing that superflares with energies $\lesssim 10^{37}$ erg are potentially capable of occurring on the Sun. The associated timescale of recurrence was found to be $\sim 20$ Myr, a value that coincided with the periodic extinction timescale of $26$ Myr deduced by some authors from the fossil record \citep{RS84}. This fact motivated us to explore the environmental and biological ramifications of $\sim 10^{36}-10^{37}$ erg superflares on Earth. In addition, we also specified the assumptions, caveats and uncertainties associated with our analysis in Sec. \ref{SSecCav}.

We concluded that a superflare of this magnitude could cause destruction of the ozone layer, thereby leading to widespread damage to ecosystems, and possibly triggering a mass extinction. In addition, the air surface temperature could rise abruptly by a considerable amount, damaging the metabolic functioning of biota because of a breakdown in thermal adaptation. We also raised the important point that small environmental perturbations could lead to far-reaching implications for ecosystems due to nonlinear processes \citep{Len08}. We also suggested that superflares may have acted in concert with geological mechanisms giving rise to extinction events that were neither wholly stochastic nor periodic. Evidence for extreme superflares may exist in the form of nitrate spikes in ice cores, anomalously high concentrations of certain cosmogenic isotopes on Earth \citep{MMN13}, and perhaps directly in the fossil extinction record.

We followed our discussion by examining some of the implications for extraterrestrial life. We inferred that present-day Mars and Venus are more susceptible to damage from superflares as they lack an intrinsic magnetic field or a thick atmosphere. We also considered exoplanets orbiting M-dwarfs and outlined why the prospects for complex life on these planets are typically lowered compared to G-type stars. A combination of factors including weak magnetic moments, close distances to the host star, extensive atmospheric stripping, and enhanced frequency of superflare events are all responsible for making environments around these stars hostile to life-as-we-know-it \citep{Mae12,LiLo17}. We also showed that, especially for planets orbiting low-mass M-dwarfs like Proxima Centauri and TRAPPIST-1, a significant fraction of the oceans can be evaporated over sub-Gyr timescales due to highly energetic superflares.  

Although superflares are likely to pose a genuine threat to human civilization, their importance has not been taken seriously in comparison to the likelihood of other astronomical catastrophes, for e.g. asteroid and comet impacts. We reviewed the literature on the economic damage wrought by superflares due to the disruption of power grids, breakdown in communications and supply chains \citep{Pul07,Hap11}. We hypothesized that the overall losses could exceed the world's current GDP for certain superflares, and that an event of this magnitude has a very high chance of transpiring during this millennium.

We completed our analysis of superflares by observing that, in certain instances, they can also lead to beneficial outcomes. During the Hadean and Eoarchean eons on Earth, when the Sun was much more active, superflares may have been an important factor in catalyzing the origin of life \citep{EIB02} and warming the planet by inducing a greenhouse effect \citep{Aira16}. With regards to the former phenomenon, superflares could have played a critical role in the synthesis of hydrogen cyanide, a vital chemical compound that is capable of giving rise to the precursors of proteins, lipids and nucleic acids under prebiotic conditions.

Superflares ought to have therefore played a major role in shaping the evolutionary history of the Earth and other habitable exoplanets. They may have constituted an essential energy source in the synthesis of prebiotic compounds, and thereby enabling abiogenesis. On the other hand, they could also have triggered quasi-periodic extinction events, although, in all probability, not the `Big Five' mass extinctions. Intriguingly, superflares might serve as putative mechanisms by which the likelihood of life on planets around M- and K-dwarfs is selectively lowered compared to G-type stars like the Sun. Thus, they provide a potential explanation as to why we, \emph{Homo sapiens}, have found ourselves dwelling on a planet orbiting the Sun instead of one that is situated in the habitable zone of an M-dwarf.\\

\acknowledgments
We thank Andrew Knoll, John Raymond, Luca Comisso, Chuanfei Dong and Russell Kulsrud for their valuable comments and suggestions concerning the paper. This work was partly supported by grants from the Breakthrough Prize Foundation for the Starshot Initiative and Harvard University's Faculty of Arts and Sciences, and by the Institute for Theory and Computation (ITC) at Harvard University.


\end{document}